\documentclass[aps,prb,twocolumn,groupedaddress]{revtex4}
\usepackage{graphics}
\usepackage{latexsym}

\begin{document}

\title{Absorbing state phase transitions with quenched disorder}

\author{Jef Hooyberghs}
\affiliation{
Departement WNI, Limburgs Universitair Centrum, 3590 Diepenbeek, Belgium}
\affiliation{
Laboratorium voor Vaste-Stoffysica en Magnetisme,
Celestijnenlaan 200D, 3001 Heverlee, Belgium}
\affiliation{VITO, Flemisch Institute for Technological Research, Boeretang 200,
2400 Mol, Belgium}
\author{Ferenc Igl\'oi}
\affiliation{
Research Institute for Solid State Physics and Optics,
H-1525 Budapest, P.O.Box 49, Hungary}
\affiliation{
Institute of Theoretical Physics,
Szeged University, H-6720 Szeged, Hungary}

\author{Carlo Vanderzande}
\affiliation{
Departement WNI, Limburgs Universitair Centrum, 3590 Diepenbeek, Belgium}
\affiliation{
Instituut voor Theoretische Fysica, Celestijnenlaan 200D, 3001
Heverlee, Belgium
}
\date{\today}
\begin{abstract}

Quenched disorder - in the sense of the Harris criterion - is generally a relevant
perturbation at an absorbing state phase transition point. Here using a strong disorder
renormalization group framework and effective numerical methods we study the properties
of random fixed points for systems in the directed percolation universality class. For strong
enough disorder the critical behavior is found to be controlled by a strong
disorder fixed point, which is isomorph with the fixed point of random quantum Ising
systems. In this fixed point dynamical correlations are logarithmically slow and the static critical
exponents are conjecturedly exact for one-dimensional systems. The renormalization group
scenario is confronted with numerical results on the random contact process in one
and two dimensions and satisfactory agreement is found. For weaker disorder the numerical
results indicate static critical exponents which vary with the strength of disorder, whereas
the dynamical correlations are compatible with two possible scenarios. Either they follow a
power-law
decay with a varying dynamical exponent, like in random quantum systems, or the dynamical
correlations are logarithmically slow even for weak disorder. For models in the parity conserving
universality class there is no strong disorder fixed point according to our renormalization
group analysis. \\

\end{abstract}

\pacs{05.70.Ln, 05.70.Jk, 64.60.Ak}

\maketitle

\newcommand{\bc}{\begin{center}}
\newcommand{\ec}{\end{center}}
\newcommand{\be}{\begin{equation}}
\newcommand{\ee}{\end{equation}}
\newcommand{\ba}{\begin{array}}
\newcommand{\ea}{\end{array}}
\newcommand{\beqn}{\begin{eqnarray}}
\newcommand{\eeqn}{\end{eqnarray}}

\section{Introduction}

Nonequilibrium many-particle systems are often described by stochastic 
models in which some degrees of freedom with fast relaxation behavior
are integrated out and are replaced by a noise term, which may represent thermal or quantum
fluctuations, chaotic motion, etc. The classification of steady states of these nonequilibrium
models is an important task and we are currently witnessing considerable theoretical progress in this field. Of particular interest are systems that in
the steady state show long-range spatial and temporal correlations, as is the case when they are
in the vicinity of a nonequilibrium phase transition point. A special type of these are
the transitions between an active phase
and an inactive one where the particles are absorbed into a state without fluctuations.
These absorbing state phase transitions play an important role in physics, chemistry and
even biology \cite{MarDic,Dic}.

Classification of absorbing state phase transitions has shown that the universality class of
directed percolation is particularly robust. It contains models with a scalar order parameter,
absence of conservation laws, and short range interactions \cite{Grass}. Well known models with a phase
transition in this universality class are the contact process \cite{Harris} and the
Ziff-Gulari-Barshad model of catalytic reactions \cite{ZGB}.
When there is a conservation law present, other universality classes can
appear, the best known of which is the parity conserving class
\cite{CardyTauber} which includes the branching-annihilating random walk
with even number of offspring\cite{BARW2,CardyTauber} and the generalized contact process with two
different absorbing states\cite{Hin}. Most recently reaction-diffusion models with multiple branching
and annihilation processes were also intensively studied\cite{multiple}.
Sandpile models with conserved energy can also be related to absorbing state phase transitions
\cite{sandpile}.

Quenched, i.e. time-independent, disorder is an inevitable feature of many real processes and could
play an important role in stochastic particle systems too. As an example, it has been argued that due to the presence of some form of disorder the directed percolation universality class
has not yet been seen in real experiments\cite{Hinexp}, such as in catalytic reactions \cite{ZGB},
in depinning transitions \cite{Barabasi}, and in the flow of granular matter \cite{Sand}
(for a review, see \cite{Hinexp}). In stochastic particle systems, disorder is represented by
position dependent reaction rates and its relevance can be expressed in a $(d+1)$-dimensional
system by a Harris-type criterion \cite{duitser},
\be
\nu_{\perp}<2/d\;.
\label{harris}
\ee
Here $\nu_{\perp}$ is the
correlation length exponent in the spatial direction of the pure system . Indeed for directed
percolation at any $d<4$ dimensions the disorder is a relevant perturbation.

The new, random fixed points, which control the critical behavior of absorbing state systems
with quenched disorder have been studied in different papers and their properties are found in
some respect to be unconventional\cite{duitser,weerDick,Janssen,BramDur,Cafiero,Webman}.
Early numerical studies by Noest\cite{duitser} on random cellular automaton models in the directed
percolation universality class in one and two dimensions have shown considerable change of
the critical exponents in comparison with the pure system's values. In more recent
studies\cite{weerDick} of the
two-dimensional contact process with dilution, logarithmically slow dynamical correlations were
found. These could be related to the results of a field-theoretical investigation by Janssen\cite{Janssen}, according to
which the renormalization group equations have only runaway solutions. Interestingly the exponents,
associated with the logarithmic time-dependence of different dynamical quantities were found
to be disorder
dependent. The absorbing phase was found to
have properties similar to that of a Griffiths phase and shows
power law behavior with non-universal exponents \cite{weerDick,BramDur,Cafiero}.
The static critical behavior of the model has been explored less,
but exponents in $d=2$ have been determined\cite{weerDick}.

In the present paper we revisit the problem of
absorbing state phase transitions in the presence of quenched disorder. In this study we make use of the formal analogy between quantum
systems and the "Hamiltonian" operator formalism\cite{Henkel,Gunther} of
stochastic processes. Although the latter operators are generally non-hermitian,
techniques developed in the study of quantum spin systems can often be successfully applied\cite{Gunther,Dickper,Enrico,HV}.

In the theory of random quantum spin systems, recently considerable progress has been
made
in understanding their low-energy (or long time), long distance behavior. In particular, the use of a
real space renormalization group (RG) method, originally introduced by Ma, Dasgupta and Hu\cite{MDH}
has lead to many new, partially exact results and has given - at the same time -
new physical insight into the problem\cite{DF}. Subsequent analytical
and numerical work has provided further important results, in particular for one-dimensional systems\cite{DF,SM,FerGrif}.
One new concept which has emerged from these studies is the existence of strong disorder fixed points
in which the disorder plays a completely dominant role. This property is manifested by the fact
that during renormalization, the distribution of the random parameters (couplings, transverse fields, etc.) broaden without limits and therefore the RG treatment becomes asymptotically exact.

Having  the close similarity between quantum systems and the Hamiltonian description of
stochastic processes, one might ask the question if the concept of strong disorder
fixed point applies
for the latter in the presence of quenched disorder. For what is perhaps the simplest stochastic process,
the random walk, the answer is positive. Sinai diffusion\cite{sinai}, which represents a particular type of random walk in a random
environment, can be interpreted as a realization of a strong disorder fixed point. Indeed RG methods have
been successfully applied to explore new exact properties of this process\cite{RGsinai}.

While the Sinai diffusion obeys detailed balance, most of the stochastic particle systems do not.
Therefore it is particularly interesting to know if quenched disorder could have a similar effect
on the latter problems, too. In this paper we are going to study this issue in detail. In
particular we investigate the applicability of the strong disorder RG scheme for models in
different (non-random) universality classes. The RG predictions are then confronted with the
results of extensive numerical calculations, which are performed
by density matrix renormalization (DMRG) and by Monte Carlo (MC) simulations. A short account
of our results has been published in a Letter\cite{hiv03}.

The structure of the paper is the following. General notations about scaling theory at absorbing state phase transitions, both at conventional and strong disorder fixed points, are given in section 2. The method of strong disorder renormalization group is explained in detail in section 3,
where it is applied to the random contact process and to its generalization with two different absorbing states. Results on random directed percolation are obtained by mapping onto a random walk. Section 4 contains a comparison of these analytical predictions with results of numerical calculations on the one- and two-dimensional contact process. Our conclusions are presented in the final section, and some technical details are given in the Appendices.

\section{Scaling at absorbing state phase transitions}

In the models that we consider here each site, $i$, of a $d$-dimensional lattice
is either vacant ($\emptyset$) or occupied by at most one particle ($A$). 
The dynamics of the model is given by a continuous time Markov process and is therefore
defined in terms of transition rates. The reactions in the system
are basically of two types: a) branching in which particles are created at empty sites
(provided one of its neighbours is occupied) occurs with rate $\lambda_i$, and
death of particles with a rate, $\mu_i$. The average number of particles at site $i$, at a given time $t$ is denoted by $\langle n_i\rangle(t)$. 
This system evolves to a stationary state in which averages become time independent. In this state, if the average value of the branching rates
compared with the average value of the death rates is sufficiently large, a finite fraction
of sites, $\rho \sim 1/L \sum_i \langle n_i\rangle>0$, is occupied and the system is in the active phase. In the opposite situation,
i.e. when the average value of the branching rates is relatively small, then $\rho=0$ and the
system is in the inactive phase. The two phases described above are separated by a phase transition
point, which is located at $\Delta=\Delta_c$, where $\Delta$ is a suitably defined control parameter.
In the homogeneous system with $\mu_i=\mu$ and $\lambda_i=\lambda$ we have $\Delta=\mu/\lambda$.

\subsection{Conventional scaling behavior}

In the vicinity of the nonequilibrium phase-transition point, where the reduced control parameter
$\delta=(\Delta-\Delta_c)/\Delta_c$ is small, generally anisotropic space-time
scaling symmetry holds. The order-parameter, $\rho(\delta,1/t,1/L)$, as a function of a
finite time-scale $t$, in a
large finite system of size $L$, scales under changing lengths by a factor $b>1$, $L'=L/b$, as
\cite{Dic,Grass}:
\be
\rho(\delta,1/t,1/L)=b^{-x}\rho(\delta b^{1/\nu_{\perp}},b^z/t,b/L)\;.
\label{scaling}
\ee
Here the critical exponents, $x$, $\nu_{\perp}$ and $z$ have the usual definitions: the correlation
length $\xi_\perp$ in the spatial (the relaxation time $t_r$ or the correlation lenght $\xi_\parallel$ in the temporal) direction diverges as $\xi_{\perp} \sim |\delta|^{-\nu_{\perp}}$
($t_r =\xi_{\parallel}\sim |\delta|^{-\nu_{\parallel}}$) and the anisotropy or dynamical exponent is
defined as $z=\nu_{\parallel}/\nu_{\perp}$. Taking the scaling parameter
$b=\delta^{-\nu_{\perp}}$ we have in the thermodynamic limit, $1/L=0$ and in the stationary state ($1/t=0$),
$\rho(\delta) \sim \delta^{\beta}$ with the order-parameter exponent $\beta=x\nu_{\perp}$.
For a surface site one defines the surface order-parameter, $\rho_s$, the scaling behavior of which
is the same as in equation (\ref{scaling}) but involves the surface scaling dimension, $x_s$,
instead of the bulk exponent $x$ so that  $\rho_s(\delta) \sim \delta^{\beta_s}$ with $\beta_s=x_s\nu_{\perp}$\cite{denen}.

The dynamical behavior of stochastic systems is related to the scaling properties of the
characteristic time, $t_r$, which in the Hamiltonian formalism is given by the inverse of
the smallest gap, $\epsilon$. In a conventional fixed point we have the relation:
\be
\epsilon(\delta,1/t,1/L)=b^{-z}\epsilon(\delta b^{1/\nu_{\perp}},b^z/t,b/L)\;.
\label{escaling}
\ee
thus in a finite system the appropriate scaling combination is $\epsilon L^z$.

The autocorrelation function, $G(\delta,1/t,1/L)=\langle n_i(t)n_i(0) \rangle$, which,
at least for a homogeneous system, is independent of the position of a particle in the bulk has the scaling behavior at the transition point:
\be
G(\delta=0,1/t,1/L)=b^{-2x}G(\delta=0,b^z/t,b/L)\;.
\label{auto}
\ee
Here, taking $b=t^{1/z}$ we obtain in the thermodynamic limit $G(t) \sim t^{-2x/z}$.

In a MC simulation one usually starts with one seed-particle in the origin of an empty lattice
and measures the survival probability at the origin, $P_s$, the total number of particles present
in the system, $N$, and the mean-square distance of the particles from the origin, $R^2$.
For the contact process it can be shown \cite{Grass,Liggett} that in the long time limit, the survival probability equals the local order parameter. This 'duality' relation also holds in the disordered case. More generally, it is expected that if there is only one absorbing state, the two quantities are expected to obey the same scaling behaviour\cite{Dic}. Thus from Eq.(\ref{scaling})
we then immediately get:
\be
P_s(\delta,1/t,1/L)=b^{-x}P_s(\delta b^{1/\nu_{\perp}},b^z/t,b/L)\;,
\label{Pscaling}
\ee
For $b=t^{1/z}$ we obtain at the critical point,
$P_s(t) \sim t^{-\theta}$ with $\theta=x/z=\beta/(\nu_{\perp} z)$. Generally we have $G(t) \sim P_s^2(t)$.
The total number of particles is proportional to the integral of the density-density
correlation function and obeys the scaling relation:
\be
N(\delta,1/t,1/L)=b^{d-2x}N(\delta b^{1/\nu_{\perp}},b^z/t,b/L)\;,
\label{Nscaling}
\ee
It therefore behaves at the critical point as $N(t)\sim t^{\eta}$, with $\eta=(d-2x)/z$.
Finally, $R^2$ scales similarly as $L^2$:
\be
R^2(\delta,1/t,1/L)=b^2 R^2(\delta b^{1/\nu_{\perp}},b^z/t,b/L)\;,
\label{Rscaling}
\ee
and at the critical point we have $R^2(t) \sim t^{2/z}$. We note that in the general situation,
i.e when several absorbing states exist, the scaling relations in
Eqs.(\ref{Pscaling}-\ref{Rscaling}) involve
a new exponent, $x'$, instead of $x$.

\subsection{Strong disorder scaling}

Strong disorder fixed points were observed so far in random quantum systems as well as in Sinai diffusion. Here we suggest that the same type of fixed point can also exist in absorbing state phase transitions and we introduce the corresponding scaling theory, which is based on two ingredients\cite{bigpaper}. First, at a strong disorder fixed point the dynamical behavior is ultraslow. The characteristic length-scale, $\xi$, is related to the logarithm of the characteristic time-scale,
$t_r$, as:
\be
\xi^{\psi} \sim \ln t_r\;.
\label{logscale}
\ee
thus the dynamical exponent, $z$, is formally infinity. Second, the average value of quantities related to the particle occupation number are dominated by so called {\it rare events} (or rare regions of a large sample). In a rare event $\langle n_i \rangle=O(1)$, thus $L$ independent, whereas in a {\it typical realization} $\langle n_i \rangle$, is generally exponentially small in $L$. The fraction of rare events, $f$, is decreasing with the size, $L$. At the critical point we can write for it the scaling transformation:
\be
f(1/L)=b^{-x}f(b/L)\;,
\label{Pr}
\ee
thus with $b=L$ we have $f \sim L^{-x}$. The average value of the particle density is indeed dominated by the rare events, thus
$\rho \sim f$, and the order-parameter at the critical point satisfies the scaling transformation:
\be
\rho(\delta,1/\ln t,1/L)=b^{-x}\rho(\delta b^{1/\nu_{\perp}},b^{\psi}/\ln t,b/L)\;.
\label{lscaling}
\ee
One can see that the static exponents, $\beta=x\nu_{\perp}$ and $\beta_s=x_s\nu_{\perp}$ are given
by the same expressions as in the case of ordinary scaling. (We shall explicitly show the construction of rare events for random directed percolation in section III.B2.)

Due to the ultraslow dynamical behavior the scaling relation (\ref{escaling}) is modified into:
\be
\ln \epsilon(\delta,1/\ln t,1/L)=b^{-\psi}\ln \epsilon(\delta b^{1/\nu_{\perp}},b^{\psi}/\ln t,b/L)\;.
\label{lescaling}
\ee
thus in a finite system the appropriate scaling combination is $(\ln \epsilon) L^{\psi}$.

To calculate the average autocorrelation function one should keep in mind that disorder in the time-direction is strictly correlated, thus in a rare event the autocorrelation function is of $O(1)$ and (almost) zero otherwise. Thus the average autocorrelation function is proportional to the fraction of rare events, $G \sim f$, and scales at the critical point:
\be
G(\delta=0,1/\ln t,1/L)=b^{-x} G(\delta=0,b^{\psi}/\ln t,b/L)\;.
\label{lauto}
\ee
Thus in the thermodynamic limit: $G(t) \sim (\ln t)^{-x/\psi}$. We can also determine the scaling behavior of other dynamical quantities such as $P_s$, $N$, and $R^2$, in the strong disorder fixed point in a similar way. It is therefore sufficient to replace in the conventional scaling relations in Eqs(\ref{Pscaling}-\ref{Rscaling}) $t$ by $\ln t$ and $z$ by $\psi$. As a consequence the time-dependence of the dynamical quantities will also be logarithmic at the critical point:
\beqn
P_s(t) \sim (\ln t)^{-\overline{\theta}}&,&\quad \overline{\theta}=x/\psi\nonumber\\
N(t) \sim (\ln t)^{\overline{\eta}}&,&\quad \overline{\eta}=(d-2x)/\psi\nonumber\\
R^2(t) \sim (\ln t)^{\overline{\sigma}}&,&\quad \overline{\sigma}=2/\psi\;.
\label{logexp}
\eeqn
This type of logarithmic time dependence has been observed by
Dickman and Moreira\cite{weerDick} by analyzing numerical data on the diluted 2d contact process, and were
interpreted as a "violation of scaling". The measured exponents $\overline{\theta}$, $\overline{\eta}$,
and $\overline{\sigma}$ were found to be dilution, i.e. disorder, dependent.

In the following, using a real space renormalization group method we i) shall give a natural
explanation of the observed logarithmic time dependence and ii) shall calculate the critical exponents, which - in the one-dimensional case - are presumably exact.

\subsection{Scaling in the Griffiths-phase}

In a disordered non-equilibrium system, which is globally in one stationary phase, say in the non-active phase with $\delta>0$, there are specific local regions of size $l_c$, in which strong fluctuations of the local rates prefer the existence of the other phase, say the active phase. These rare regions, which are localized and have an exponentionally small probability of occurance, $p(l_c) \sim \exp(-\alpha l_c)$, contribute to an exponentially large relaxation time\cite{note_gr}, $t_r \sim \exp(\sigma l_c)$. (In a d-dimensional system $l_c$ should be replaced by $l_c^d$.) Then the distribution of large relaxation times has an algebraic tail: $p(t_r) \sim t_r^{-1/z'-1}$, with $1/z'=\beta/\sigma$ and the average autocorrelation function:
\be
G(t) \sim \int {\rm d} t_r p(t_r) \exp(-t/t_r) \sim t^{-1/z'}\;
\label{auto_g}
\ee
also decays algebraically. Consequently in this so called Griffiths phase\cite{griffiths} dynamical correlations are quasi-long-ranged, whereas spatial correlations are short ranged. The dynamical exponent is a continuously varying function of the distance from the critical point, $z'=z'(\delta)$.

This result can be incorporated into a scaling theory as follows. Since a rare event, which brings the dominant contribution to the average autocorrelation function is localized, its probability of occurance is inversely proportional with the size of the system, $L$. Consequently the average autocorrelation function obeys the scaling law:
\be
G(\delta,1/t,1/L)=b^{-1} G(\delta,b^{z'}/t,b/L)\;,
\label{auto_g1}
\ee
and with $b=t^{1/z'}$ we recover in the thermodynamic limit the relation in Eq.(\ref{auto_g}). Finally, scaling of the lowest gaps follow the rule:
\be
\epsilon(\delta,1/t,1/L)=b^{-z'}\epsilon(\delta,b^{z'}/t,b/L)\;.
\label{gescaling}
\ee
thus in a finite system the appropriate scaling combination is $\epsilon L^{z'}$. We note that in the Griffiths phase the power-law singularities are often supplemented by logarithmic corrections\cite{huse,i02}, which are related to the fact that the size of the rare event grows logarithmically, $l_c \sim \ln L$, since $p(l_c) \sim 1/L$.

\section{The renormalization group framework}

In this section we apply a real space renormalization group method for random stochastic
particle systems, which is a variant of the Ma-Dasgupta-Hu method originally developed to study random quantum spin chains\cite{MDH}. The essence of the method can be summarized in the following points.

i) Start with the initial distribution of the random reaction rates, $P_{\rm in}(\mu)$ and
$R_{\rm in}(\lambda)$ and sort the rates in descending order. The largest rate,
i.e. the fastest process, sets the energy scale,
$\Omega$, in the problem.

ii) Integrate out the fastest local process, i.e. eliminate the rate, $\Omega$. This amounts to decimate out one site of the lattice or to replace a pair of sites with a new
effective one. Renormalised couplings are then determined using a second order perturbation
calculation. It is interesting to remark that for stochastic systems this step corresponds
to a fast rate expansion as discussed in section 4.3 of reference 21. 

iii) Iterate the decimation process. This will result in a reduction of the energy-scale, $\Omega$,
and a modification of the distribution of the (effective) rates: $P(\mu,\Omega)$ and
$R(\lambda,\Omega)$.

iv) At the fixed point of the transformation (which is at $\Omega=\Omega^*=0$) the distributions of the rates become singular and from these singularities the value of the critical exponents are calculated.

In the following we construct and solve the RG equations explicitly for the random contact process.

\subsection{The random contact process}

\subsubsection{The Hamiltonian formalism}

In the contact process each site of the lattice can be either vacant ($\emptyset$) or occupied by at most one particle ($A$), and thus can be characterized by an Ising-spin variable, $\sigma_i=1$ for $\emptyset$ and $\sigma_i=-1$ for $A$. The state of the system is then given
by the vector ${\bf P}({\bf \sigma},t)$ 
which gives the probability that the system is in the state ${\bf \sigma}=\{\dots ,\sigma_i, \dots\}$ at time $t$.  A particle can be created at an empty site $i$ with
a rate $p \hat{\lambda_i}/p_0$, where $p$ ($p_0$) is the number of occupied neighbors (the coordination
number of the lattice) and at an occupied site the particle is annihilated with a rate $\mu_i$. The time evolution is governed by a master equation, which
can be written into the form:
\be
\frac{{\rm d} {\bf P}({\bf \sigma})}{{\rm d} t}=-H_{CP} {\bf P}({\bf \sigma})\;.
\label{master}
\ee
Here the generator $H_{CP}$ of the Markov process is given by:
\beqn
H_{CP} = \sum_i
\mu_{i} M_{i}
+ \sum_{\langle ij \rangle} \frac{\hat{\lambda}_{i}}{p_0} (n_{i} Q_{j} + Q_{i} n_{j})
\label{hamilton}
\eeqn
in terms of the matrices:
\begin{eqnarray*}
M = \left(\begin{array}{rr}
0 & -1\\
0 & 1\end{array}\right),
n = \left(\begin{array}{rr}
0 & 0\\
0 & 1\end{array}\right),
Q = \left(\begin{array}{rr}
1 & 0\\
-1 & 0\end{array}\right)
\end{eqnarray*}
and $\langle ij \rangle$ stands for nearest neighbors. It is well known\cite{Gunther} that the
steady state probability distribution of a stochastic process coincides with the
ground state of its generator (sometimes also called quantum Hamiltonian of
the stochastic process) while relaxation properties can be determined
from its low lying spectrum.

For non-random couplings there is a non-equilibrium phase transition in the contact process
which belongs to the universality class of directed percolation\cite{Dic,Grass}. In
one dimension it is at $(\mu/\hat{\lambda})_c=0.3032$ and the critical exponents are given by:
$\beta=0.2765$, $\beta_s=0.7337$, $\nu_{\perp}=1.097$ and $z=1.581$. In two dimensions
$(\mu/\hat{\lambda})_c=0.6065$ and $\beta=0.584$, $\beta_s=1.03$, $\nu_{\perp}=0.734$ and $z=1.76$.
In the following we often use the variable $\lambda=\hat{\lambda} / p_0$ to characterize the creation
rate.

\subsubsection{Decimation rules of the random process}

For the random contact process the transition rates $\mu_i$ and $\lambda_i$ are independent
and identically distributed variables and, as described in the previous section, the largest one $\Omega={\rm max}(\{\lambda_i\},
\{\mu_i\})$ sets the energy scale in the system. Thus, the largest rate can be either one of the
death rates $\mu_i$ or be one of the branching rates $\lambda_i$. For each a different way of decimation should be used.

\bigskip
i) \underline{The largest term is a branching rate: $\Omega=\lambda_2$}

If the largest rate is a branching, say $\lambda_2$, which connects sites $i=2$ and $j=3$, the two-site cluster $(2,3)$ spends
most of the time in the configurations $AA$ or $\emptyset\emptyset$ and can be rarely found in one of the
other two configurations, $A \emptyset$ and $\emptyset A$. Consequently for large times the two sites behave
as a cluster with a moment of $\tilde{m}=2$ and with an effective death rate, $\tilde{\mu}_2$,
which can be calculated from the energy spectrum of the two-site Hamiltonian, $H_{CP}^{23}$. As shown
in the Appendix A the two lowest energy levels of the cluster are separated from the two others by a distance of
$\lambda_2$, therefore in a good approximation just the two lowest levels can be retained. The effective
death rate, $\tilde{\mu}$, is given from the value of the lowest gap as:
\be
\tilde{\mu}=\frac{2 \mu_2 \mu_3}{\lambda_2}\;,
\label{mutilde}
\ee
where $\mu_2$ and $\mu_3$ are the original death rates at sites $i=2$ and $j=3$,
respectively. The renormalization equation
in Eq.(\ref{mutilde}) should be extended by the renormalization of moments (i.e. the number of original sites in the cluster):
\be
\tilde{m}=m_2 + m_3\;,
\label{moments}
\ee
where in the initial situation $m_2=m_3=1$.

The renormalized value of the death rate can also be obtained with the following reasoning. Let us start with the original representation, when the two-site cluster is in the occupied state, $AA$. In the effective decay process first the particle
at site $2$  should decay (with rate $\mu_{2}$), which is then
followed by the decay of the particle at $3$. This second
process has a very low probability of $\mu_{3}/(\lambda_{2}+\mu_{3})$.
Since the same processes can also occur with the role of $2$ and $3$ interchanged,
we find that for $\lambda_{2} \gg \mu_{2}, \mu_{3}$ the effective
decay rate is given in Eq.(\ref{mutilde}).

\bigskip
ii) \underline{The largest term is a death rate: $\Omega=\mu_2$}

In this case the site (2) is almost always empty, $\emptyset$, therefore it does not
contribute to the fractal properties of the $A$ cluster and can be decimated out. The
effective branching rate, $\tilde{\lambda}$ between the remaining sites $1$ and $3$ can
be obtained by studying the eigenvalues of the three-site Hamiltonian, $H_{CP}^{123}$. As
shown in the Appendix A out of the eight eigenvalues there are four of $O(\mu_2)$, which are
discarded. The remaining lowest four levels are identified as the spectrum of a two-site
cluster with an effective branching rate:
\be
\tilde{\lambda}=\frac{\lambda_2 \lambda_3}{\mu_2}\;.
\label{lambdatilde}
\ee
As in the previous case one can obtain the renormalized value of the branching rate with a simple
reasoning, which works as follows. Let us have the configuration of the three-site cluster in the original representation as $A\emptyset\emptyset$. The effective branching rate between sites $1$ and $3$ is generated
by a virtual process, in which first a particle is created at site $2$ (rate
$\lambda_{2}$), and then one at site $3$ (probability $\lambda_{3}/(
\lambda_{3} + \mu_{2})$). Hence, we get for very strong disorder the branching rate given in
Eq.(\ref{lambdatilde}).

The renormalization equations in Eqs.(\ref{mutilde}) and (\ref{lambdatilde}) can be transformed
into a symmetric form in terms of the variable,
$J= \lambda/\kappa=\hat{\lambda}/(p_0\kappa)$ with $\kappa=\sqrt{2}$ as
\begin{eqnarray}
\tilde{\mu} = \kappa \frac{\mu \mu'}{J},
\quad \tilde{m}=m+m',
\ \ \ \ \ \tilde{J} = \kappa
\frac{J J'}{\mu}\;.
\label{RG}
\end{eqnarray}
We can see from Eq.(\ref{RG}) that for weak disorder the generated new rates can be occasionally larger then the decimated ones, thus in these steps the energy-scale does not lower. For strong enough disorder, however, these non-monotonic steps are expected to be so rare that they do not influence the behavior of the RG flow. With this assumption we analyze in the following the properties of the RG equations in one dimension. The results are then confronted with numerical calculations in section IV.

\subsubsection{Renormalization in one dimension}

In one dimension the topology of the lattice does not change under renormalization which makes it possible to treat the problem analytically.
First we note that after a repeated use of the transformations in Eq.(\ref{RG}) the
generated branching (death) rates are in the form of a ratio of products of original
branching (death) rates and original death (branching) rates. The control
parameter, $\delta$, is defined as:
\begin{equation}
\delta=\frac{[\ln \mu]_{\rm av}- [\ln J]_{\rm av}}
{{\rm var}[\ln \mu]+{\rm var}[\ln J]}\;,
\label{contr}
\end{equation}
and at the fixed point, $\delta=0$, which follows from duality of the RG equations
in Eq.(\ref{RG}) (here $[.]_{\rm av}$ denotes the average over disorder).

At a given energy scale, $\Omega$, we have the distribution function of the
death rates, $P(\mu,\Omega)$, and that of the branching rates, $R(J,\Omega)$.
Changing the energy scale, $\Omega \to \Omega - {\rm d} \Omega$, amounts
to eliminate a fraction of ${\rm d} \Omega [ P(\Omega,\Omega)+R(\Omega,\Omega)]$
sites. The distribution of the branching rates changes as:
\begin{widetext}
\begin{eqnarray}
R(J,\Omega-{\rm d} \Omega)=\left\{R(J,\Omega)+{\rm d} \Omega P(\Omega,\Omega)
\int_0^{\Omega} {\rm d} J_1 \int_0^{\Omega} {\rm d} J_3 R(J_1,\Omega)R(J_3,\Omega)
\times \right.
\nonumber\\
\left. \left[ \delta\left(J-\frac{\kappa J_1 J_3 }{\Omega }\right)-\delta(J-J_1)-\delta(J-J_3)\right]
\right\}\left\{1-{\rm d} \Omega [ P(\Omega,\Omega)+R(\Omega,\Omega)] \right\}^{-1}
\;.
\label{Jdistr}
\end{eqnarray}
\end{widetext}
Here in the r.h.s. the three delta functions represent the one generated new branching rate and the
two decimated branching rates during one RG step and the second factor ensures normalization.\cite{note1}. A similar
equation is obtained for the distribution of the death rates. From the duality
of the RG equations, it follows that one should
only make the interchange $\mu \leftrightarrow J$ and $P \leftrightarrow R$.

Expanding $R(J, \Omega-{\rm d}\Omega)$ one arrives to the integro-differential
equation:
\begin{eqnarray}
\frac{{\rm d} R}{{\rm d} \Omega}=R(J,\Omega)\left[P(\Omega,\Omega)-R(\Omega,\Omega)\right]
\nonumber\\
-P(\Omega,\Omega) \int_{J/\kappa }^\Omega {\rm d} J' R(J',\Omega)
R\left(\frac{J}{J'\kappa}\Omega ,\Omega\right) \frac{\Omega}{J'\kappa}
\;,
\label{Rdiff}
\end{eqnarray}
and similarly one obtains for the distribution $P(\mu,\Omega)$:
\begin{eqnarray}
\frac{{\rm d} P}{{\rm d} \Omega}=P(\mu,\Omega)\left[R(\Omega,\Omega)-P(\Omega,\Omega)\right]
\nonumber\\
-R(\Omega,\Omega) \int_{\mu/\kappa }^\Omega {\rm d} \mu' P(\mu',\Omega)
P\left(\frac{\mu}{\mu'\kappa}\Omega ,\Omega\right) \frac{\Omega}{\mu'\kappa}
\;.
\label{Pdiff}
\end{eqnarray}
Solution of these equations can be obtained analytically at the fixed point, $\Omega=0$, at $\delta=0$, in which the distributions $R(J,\Omega)$ and $P(\mu,\Omega)$ are asymptotically identical. The calculations can be found in the Appendix B.
According to these results the appropriate scaling variable is in logarithmic
form $\eta=-(\ln \Omega
-\ln J)/\ln \Omega=-(\ln \Omega -\ln \mu)/\ln \Omega$, and its distribution is given from Eqs.(\ref{Rsol}) and (\ref{sol0}) as
\begin{equation}
p(\eta){\rm d} \eta = \exp(-\eta){\rm d} \eta \;.
\label{soleta}
\end{equation}
The distribution in terms of the original variables, $J$ (and $\mu$) in Eq.(\ref{Psol})
is given by
\begin{equation}
R(J,\Omega)=\tilde{R}\left(\frac{\Omega}{J}\right)^{1-\tilde{R}\Omega}, \quad
\tilde{R} \Omega=\frac{1}{\ln(\Omega_0/\Omega)},
\label{Rsol+}\
\end{equation}
where $\Omega_0$ is a reference energy scale, and
the distribution becomes singular at the fixed point, as $\Omega \to 0$.
Due to this singularity the decimation transformation in Eq.(\ref{RG}) becomes exact at the fixed
point. This can be shown by calculating the probability that one of the neighboring
death rates, beside the largest branching rate with $J=\Omega$, has a value of $\mu>\alpha\Omega$,
with $\alpha<1$:
\begin{eqnarray}
P(\alpha) &\simeq& \int_{\alpha\Omega}^{\Omega} P(J,\Omega) {\rm d} J=\tilde{R} \Omega
\int_{\alpha}^1 x^{-1+\tilde{R} \Omega}{\rm d} x \nonumber \\
&\approx &
\tilde{R} \Omega \ln(1/\alpha)\;,
\label{Palpha}
\end{eqnarray}
which indeed goes to zero as the iteration proceeds, since $\tilde{R} \Omega \to 0$. Consequently
the RG transformation becomes asymptotically exact and the singularities, calculated
by this method at the critical point, are also very probably exact.

We start to determine the relation between the energy scale, $\Omega$, and the
length scale, $L_{\Omega}$, by studying the fraction of non-decimated sites,
$n_{\Omega}$. When the energy
scale is decreased by an amount of ${\rm d}\Omega$ a fraction of sites.
${\rm d} n_{\Omega}= n_{\Omega}[\tilde{P}(\Omega)+\tilde{R}(\Omega)]$, is decimated
out, so that we obtain the differential equation:
\begin{equation}
\frac{{\rm d} n_{\Omega}}{{\rm d} \Omega}= n_{\Omega}[\tilde{P}(\Omega)+\tilde{R}(\Omega)]
\;,
\label{dnomega}
\end{equation}
which can be rewritten as
\begin{equation}
-\frac{{\rm d} \ln n_{\Omega}}{{\rm d} \ln \Omega}= -\Omega [\tilde{P}(\Omega)+\tilde{R}(\Omega)]=
-2y(\Omega)
\;.
\label{dnoGamma}
\end{equation}
Using the solution to $y(\Omega)$ in Eq.(\ref{sol0}) one can integrate Eq.(\ref{dnoGamma})
with the result:
\begin{equation}
n_{\Omega}= \left[ 1+y_0 \ln \frac{\Omega_0}{\Omega} \right]^{-2}
\sim \left[ \ln \frac{\Omega_0} {\Omega}\right]^{-2},\quad \delta=0
\;.
\label{nomega0}
\end{equation}
Thus from Eq.(\ref{nomega0}) we get
for the typical distance between remaining spins, $L_{\Omega}$, as:
\begin{equation}
L_{\Omega}\sim \frac{1}{n_{\Omega}} \sim \left[ \ln \frac{\Omega_0}{\Omega}\right]^2,
\quad \delta=0
\;,
\label{lomega0}
\end{equation}
This is equivalent with a logarithmic dynamical scaling as written in Eq.(\ref{logscale})
with an exponent, $\psi=1/2$.

In order to calculate the singularity of other quantities, such as the correlation length and the order-parameter, at the fixed point we should study scaling of the lengths and the cluster moments. As we have shown in the Appendix C, at the fixed point these calculations are equivalent to that for the random transverse Ising spin chain. Therefore here we quote only the results. For a detailed derivation we refer to the original literature \cite{DF,i02}.

We remind first that a renormalized site is composed from parts of the original lattice and the renormalized length is given by the sum of the lengths in the original lattice.
In the paramagnetic phase the average length of sites approach a finite value, $\xi_{\perp}$, during renormalization as $\Omega \to 0$. In the vicinity of the fixed point the RG-equations lead to a singularity:
\be
\xi_{\perp} \sim |\delta|^{-\nu_{\perp}},\quad \nu_{\perp}=2\;.
\label{nu_perp}
\ee

To study scaling of the order-parameter one should investigate the average cluster moment, which at the critical point behaves as:
\begin{equation}
\overline{m}= \overline{m}_0 \left[ \ln
\left(\frac{\Omega_0}{\Omega}\right)\right]^{\Phi},\quad
\Phi=\frac{1}{\tau}=\frac{1+\sqrt{5}}{2}\;.
\label{Phi}
\end{equation}
Then the order-parameter can be calculated as $\rho=\overline{m}/L_{\Omega}$, which behaves asymptotically as
\begin{equation}
\rho \sim L^{-x}\;,
\label{x_m1}
\end{equation}
and the scaling dimension, $x$, is given by Eqs.(\ref{lomega0}) and (\ref{Phi}) as
\begin{equation}
x=\frac{2-\Phi}{2}\;.
\label{x_m}
\end{equation}
Finally, scaling of the surface order-parameter is related to the average cluster moment of the surface site, $\overline{m_s}$. This is naturally smaller than for a bulk site, since the surface moment can accumulate from original sites only in one direction. According to an analysis of the RG results we have:
\begin{equation}
\overline{m_s}= \overline{m_s}_0 \left[ \ln
\left(\frac{\Omega_0}{\Omega}\right)\right]\;.
\label{m_s}
\end{equation}
thus
\begin{equation}
\rho_s \sim L^{-x_s}, \quad x_s=1/2\;.
\label{x_ms}
\end{equation}
\subsubsection{Renormalization in two dimensions}

In higher dimensions the topology of the lattice changes under renormalization: contacts and therefore reactions are generated between remote sites, too. However, the renormalization does not introduce new types of reactions. Therefore the renormalization process, which is summarized in the decimation equations in Eq.(\ref{RG}) can be implemented numerically. As in one dimension for weak disorder the generated new couplings are frequently larger than the decimated ones, therefore the RG scheme does not work and scaling in the random system is most probably conventional, as described in section 2A. For stronger disorder, however, the situation could change and strong disorder scaling could set in. If the critical behavior of the system is indeed attracted by a strong disorder fixed point then, as in one dimension, the value of a finite pre-factor $\kappa>0$
in Eq.(\ref{RG}) does not matter. Numerical renormalization group calculations for the random transverse-field Ising model, for which $\kappa=1$, have shown the existence of a strong disorder fixed point in two-dimensions\cite{2dRTIM,2dRTIM1}. The numerically observed critical exponents are: $x=1.0$, $\nu_{\perp}=1.07$ and $\psi=.42$ \ \cite{2dRTIM}. Karevski {\it et al.}  \cite{2dRTIM1} use  a somewhat different numerical technique and find:
$x=0.97$, $\nu_{\perp}=1.25$ and $\psi=.5$. In the light of the above arguments we expect also for the two-dimensional random contact process strong disorder scaling with the above exponents in Eq.(\ref{lscaling}) if the strength of disorder is sufficiently large. To verify this scenario we shall reanalyze the numerical results of Moreira and Dickman \cite{weerDick} in section IV.B.

\subsection{Random directed percolation}

Directed percolation can be viewed as an anisotropic variant of percolation, in which the possible path of occupied sites follows a given preferential direction. Equivalently, directed percolation can be interpreted as a dynamical process, in which the spreading of a non-conserved agent is studied. In the random version of the problem the occupation probabilities are random variables, which are strictly correlated along the preferential (time) direction, corresponding to random reaction rates in the dynamical interpretation.

\subsubsection{Random Reggeon field theory}

The field theory, which is expected to describe the critical behavior of directed percolation is the Reggeon field theory. In $(1+1)$-dimension in the Hamiltonian limit the time-evolution of the process is governed by the Hamiltonian\cite{QRFT} $H_{RF}=\sum_i H_i$ with:
\be
H_i= -h_i \sigma_i^x -  \frac{g_i}{2}\left[(1-2\sigma_i^+) (1-2\sigma_{i+1}^+)- \sigma_i^z \sigma_{i+1}^z \right]
\label{H_RF}
\ee
where $\sigma_i^{x,y,z}$ are Pauli matrices at site, $i$, and $\sigma_i^+=(\sigma_i^x+i\sigma_i^y)/2$. The structure of $H_{RF}$ is similar to the Hamiltonian of the random transverse-field Ising model in Eq.(\ref{hamilton}): it consists of an interaction term (which is non-hermitian in $H_{RF}$) and a transverse field term. The order in the system is measured by the asymptotic limit of the autocorrelation function, $G(t)=[\langle 0|\sigma_i^z(t)\sigma_i^z(t)|0\rangle]_{\rm av}$: it is zero in the paramagnetic phase and finite in the ordered phase.

The renormalization of the random Reggeon-field theory can be made in a way that is
completely similar to that for the random contact process (or the random transverse-field Ising model). A very strong coupling, $g_2=\Omega$, will result in a two-site cluster in a renormalized field which is given by:
\be
\tilde{h}=\frac{2 h_2 h_3}{g_2}\;.
\label{htilde}
\ee
On the other hand a site on which a very strong field, $h_2=\Omega$, acts is decimated out and a new coupling is generated between remaining sites, as:
\be
\tilde{g}=\frac{g_2 g_3}{h_2}\;.
\label{gtilde}
\ee
Comparing the decimation equations in Eqs.(\ref{htilde}) and (\ref{gtilde}) with those for the random contact process in Eqs.(\ref{mutilde}) and (\ref{lambdatilde}) we can see that they are equivalent. This is not surprising, since the two Hamilton operators in Eqs.(\ref{hamilton}) and (\ref{H_RF}) are related through a unitary transformation\cite{Grass}, if $\mu_i \leftrightarrow h_i$ and $\lambda_i \leftrightarrow g_i$. Consequently the singular behavior of the random contact process and the random Reggeon field theory are equivalent.

Next, we consider the geometrical interpretation of directed percolation and study its critical behavior in the very strong disorder limit.

\subsubsection{Strong disorder: mapping to random walks}

\begin{figure}
\resizebox{8cm}{8cm}{\includegraphics{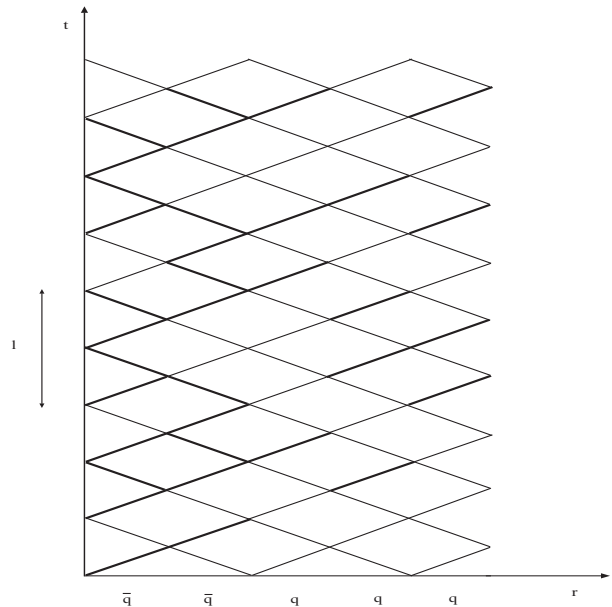}}
\caption{Directed percolation on the square lattice with random bond occupation
probabilities, which are perfectly correlated along the vertical diagonal direction.
Occupied bonds are drawn in bold.
In the two columns on the left bonds are occupied with a large probability
$\overline{q}$, while in the other three columns this probability is small and equals $q$.
The typical length-scale in the problem is $l \sim 1/q$. The direction of time, $t$, is
also indicated which gives a reinterpretation of directed
percolation in terms of a reaction-diffusion model. \label{fig.1}}
\end{figure}

Here we consider directed percolation on the square lattice with random occupation probabilities which are, however, strictly correlated in the same layer, as shown in Fig. \ref{fig.1}.

For simplicity, we use a bimodal distribution, the occupation probability in the $i$-th layer can be either $p_i=q$ with probability $\pi$ or $p_i=1-q=\overline{q}$, with probability $\overline{\pi}$ (we take $q \leq \overline{q}$). For isotropic percolation the critical point is located at $\pi=\overline{\pi}=1/2$, which follows from self-duality\cite{ji02}. Since directed  clusters generally contain less bonds than the isotropic ones, for directed percolation at the critical point $\pi<\overline{\pi}$. In the strong disorder limit we take $q \to 0$ in which case at the critical point $(\pi/\overline{\pi})_c \to 1$ as we shall show below. The effect of the same type of disorder for isotropic percolation has already been considered by one of us \cite{ji02}. The reasonings in the two problems are very similar and they lead to identical critical properties. Therefore, here we just briefly describe the method and emphasize the relations and differences between the directed and isotropic problems. For more details we refer to the original reference \cite{ji02}.

The characteristic structure of the occupied bonds is very different for the two types of layers. In layers with a large probability, $p_i=\overline{q}$, almost all bonds are occupied, between two unoccupied bonds there is a characteristic distance, $l \sim 1/q$. On the other hand in layers with a small probability, $p_i=q$, there are only very few occupied bonds, two occupied bonds are separated by the same characteristic distance, $l \sim 1/q$. Note the duality: for layers with $p_i=q$ the non-occupied (say white) bonds play the same role as the occupied (say black) bonds for layers with $p_i=\overline{q}$.

Now let us consider a system consisting of $r=1,2,\dots,L$ layers with free boundary conditions, with the seed at $r=0$ and determine the surface order-parameter, $\rho_s(L)$, which is given by the probability that the percolating cluster extends up to the other surface at $r=L$. To calculate $\rho_s(L)$ we consider strips of width $k=1,2,\dots L$ and estimate, $n(k)$, the typical number of sites in the $k$-the (i.e. surface) layer which are connected to the seed. It is evident from the duality properties that for $n(k) \gg 1$ the probability that the seed is connected to the $k$-th layer through {\it white} bonds is $p_w(k) \sim 1/n(k)$, since typically one out of $n(k)$ sites has this property. We show by induction that $n(k)$ is either zero or given by $n(k) \sim q^{-X(k)}$, where $X(k) >0$ is the number of layers with a probability $\overline{q}$ minus the number of layers with a probability $q$. First, our statement is trivially true for $k=1$. Here $n(1) \sim 1/2q$ (and thus $p_w(k) \sim q$), for $p_1=\overline{q}$ and $n(1)=0$, for $p_1=q$. Evidently, if in a given sample $n(k)=0$ for some $k<L$, then $n(k')=0$ for any $k'>k$ and the surface order-parameter is zero. To complete our proof, in the second step we show that, if $X(k) \ge 1$, then $X(k+1)=X(k) \pm 1$, where the upper (lower) sign stands for a probability $p_{k+1}=\overline{q}$  ($p_{k+1}=q$). The proof of this statement follows from the fact, that for $p_{k+1}=q$ the number of black bonds in the cluster at layer $k$ is reduced by a factor of $\sim q$, whereas for $p_{k+1}=\overline{q}$ the probability $p_w(k)$ is reduced by a same factor $\sim q$. Here, when $X(k)=1$ and $p_{k+1}=q$, thus $X(k+1)=0$ the cluster is considered to be terminated at this point, thus $\rho_s(L)=0$.

To calculate the average value of the surface order-parameter we use a random walk picture, which has already been applied in the isotropic problem \cite{ji02}. To a given sample with a given probability distribution we assign a random walk which starts at
position $X_0=0$ and makes its $i$-th step upwards (downwards) for $p_i=\overline{q}$
($p_i=q$). The position of the walker in the $k$-th step is just $X_k$. The existence of finite surface order-parameter in the given sample is then formulated by the condition:
$n(k) > 0$ for $k=1,2,\dots,L$, thus $X_k > 0$ for $k=1,2,\dots,L$ so that the random walk has a surviving character. The {\it average value} of the surface order-parameter is given by the fraction of samples with finite surface order, which is just the survival probability of the random walk: $[\rho_s]_{\rm av}(L) \sim P_s(L) \sim L^{-1/2}$. From this relation the value of the surface order-parameter scaling dimension, $x_s=1/2$ follows,
which is the same as for the random contact process in the strong disorder fixed point in Eq.(\ref{x_ms}). Note that the rare events introduced in section II.B in this case are the samples with a surviving characteristics in the probability distribution. Furthermore in the critical situation, when the average surface order-parameter decays as a power with $L$ the probabilities, $\pi=\overline{\pi}$, as announced before.

Other exponents can be deduced in analogous way as for isotropic percolation. The typical temporal extent of the percolating cluster is given by the typical number of connected sites in a layer, $\xi_{\parallel} \sim n_{typ}(L) \sim q^{-X_{typ}}$,
where the typical excursion of a (surviving) random walk in $L$ steps is $X_{typ} \sim L^{1/2}$. Thus we have the logarithmic scaling relation in Eq.(\ref{logscale}) with $\psi=1/2$, as for the random contact process. To calculate the bulk order-parameter the seed is put in the middle of the lattice and in a given sample $\rho(L)=O(1)$ if the percolating cluster has an extent of $L$. In the language of random walks this property is related to the so called average persistence\cite{ri99}. From this, scaling of the average (bulk) order-parameter is given by $[\rho]_{\rm av}(L) \sim L^{-x}$, with $x=(3-\sqrt{5})/4$, as for the random contact process in Eq.(\ref{x_m}). Finally, to determine the scaling exponent of the order-parameter in the vicinity of the critical point one should consider random walks with finite bias, $\delta_w \ne 0$, towards the absorbing walk and $\delta \sim \delta_w$. From the scaling of the surviving probability of biased walks\cite{bigpaper} one obtains for the correlation length, $\xi_{\perp} \sim |\delta|^{-2}$. Thus we recover the same exponent, $\nu_{\perp}=2$, as for the random contact process in Eq.(\ref{nu_perp}).

Hence we can conclude that the critical properties of random directed percolation can be deduced from a random walk mapping in the strong disorder limit. The critical behavior is the same as for random isotropic percolation and corresponds to the RG results obtained for the random contact process in the strong disorder fixed point.

\subsection{The generalized contact process with disorder}

So far we have considered different variants of absorbing state phase transitions which, in the absence of disorder, all belong to the directed percolation universality class. We observed, that in the presence of strong enough quenched disorder all these processes show strong disorder scaling behavior with identical critical exponents. In this section we consider other processes, which are not in the directed percolation universality class.

For a renormalization group treatment, one of the most convenient models is the generalized contact process with several different absorbing states introduced by Hinrichsen\cite{Hin}. In this model a site can be occupied by a particle, $A$, or can be in one of $n$ empty states $\emptyset_1,\emptyset_2, \dots \emptyset_n$. Furthermore, besides the rules known for the ordinary contact process, there is a competition between the different type of empty states. At the border of clusters with different type of empty states particles can be created. As a consequence, in these models, with increasing $n$ there is a preference for the active phase and the phase transitions is found to be in universality classes different from the directed percolation one. For $n=2$, the model was shown to be in the parity conserving class\cite{Hin}, whereas for $n \ge 3$ the model is always active\cite{hcv01}.

Here we consider in particular the effect of disorder on the model with $n=2$.
For this case, the following processes are allowed:
\beqn
AA \to A \emptyset_1,A \emptyset_2, \emptyset_1 A, \emptyset_2 A \quad &{\rm rate}& \quad \mu_i/2
\nonumber\\
A \emptyset_1, \emptyset_1 A \to \emptyset_1 \emptyset_1; \quad A \emptyset_2, \emptyset_2 A \to \emptyset_2 \emptyset_2 \quad &{\rm rate}& \quad \mu_i
\nonumber\\
A \emptyset_1, A \emptyset_2, \emptyset_1 A, \emptyset_2 A \to AA \quad &{\rm rate}& \quad \lambda_i
\nonumber\\
\emptyset_1 \emptyset_2 \to \emptyset_1 A, A \emptyset_2; \quad \emptyset_2 \emptyset_1 \to \emptyset_2 A, A \emptyset_1 \quad &{\rm rate}& \quad \lambda_i\;
\label{rate_gcp}
\eeqn
For non-random couplings the phase-transition in one dimension is located at $\mu/\lambda=1.592$ and the critical exponents are $\nu_{\perp}=1.82$, $z=1.75$, $\beta=0.91$ consistent with those of the parity conserving universality class. Note that the Harris-type criterion in Eq.(\ref{harris}) with the above $\nu_{\perp}$ predicts a relevant perturbation for quenched disorder.

Next, for the random system we try to apply the renormalization group method along the lines used for the random contact process in section III.A. We start with the decimation scheme and consider the situation when  the largest rate is a branching rate, say $\lambda_2=\Omega$. As for the random contact process we take the two-site Hamiltonian, which contains now $9$ states, since each site could be in three (one active and two different inactive) states, and calculate the lowest three eigenstates along the lines presented in the Appendix A for the random contact process. Among the three lowest states, which are kept after decimation, two have eigenvalues, $0$, and the third has an energy $\epsilon_1^{23} \simeq 2 \mu_2 \mu_3/\lambda_2$. Thus after decimation we have an effective two-site cluster in the presence of a renormalized death rate, which is given just in the same form as in the random contact process, see Eq.(\ref{mutilde}). The value of the effective death rate can be also obtained by a similar argument as for the random contact process. In order to calculate the decay from $AA \to \emptyset_1 \emptyset_1, \emptyset_2\emptyset_2$ one should consider two second-order virtual processes: $AA \to A \emptyset_1 \to \emptyset_1 \emptyset_1$ and $AA \to A \emptyset_2 \to \emptyset_2 \emptyset_2$, which leads to Eq.(\ref{mutilde}) by taking into account the definition of the rates in Eq.(\ref{rate_gcp}).

For the case of a strong elimination rate, $\mu_2=\Omega$, the three site Hamiltonian contains $27$ eigenstates, which are divided into $9$ orthogonal sectors, each of which have $3$ states. The highest levels of each sector have an energy of $O(\mu_2)$ and thus can be discarded during decimation. The remaining $18$ states, however, are all in the same order of magnitude. Three sectors have ground state energy zero and first excitation energy,
$\lambda_2+\lambda_3$. Two other sectors have the lowest energies:
\be
\epsilon_2^{123}=\left[[7\lambda_2+4\lambda_3) \pm \sqrt{(7\lambda_2+4\lambda_3)^2-16 \lambda_2 \lambda_3}\right]/8\;,
\label{e_2}
\ee
and in another two we should exchange in Eg.(\ref{e_2}) $\lambda_2$ and $\lambda_3$. Finally the last two sectors are also degenerate with the lowest eigenvalues:
\be
\epsilon_3^{123}=\left[3(\lambda_2+\lambda_3) \pm \sqrt{9(\lambda_2+\lambda_3)^2-32 \lambda_2 \lambda_3}\right]/4\;.
\label{e_3}
\ee
Consequently the decimation does not work out here for a large death rate, since i) one can only discard $9$ out of the $27$ cell states, so that the remaining states can not be assigned to one renormalized site and ii) the remaining energy levels are of the same order as the original rates, thus the energy scale is not lowered during these steps.
These features of the decimation can be seen by the following argument, too. With a large death rate, $\mu_2=\Omega$, the site $i=2$ is almost always inactive, so that it is either in $\emptyset_1$ or $\emptyset_2$ most of the time. Suppose we decimate this site  and calculate the effective rate for a process in which in
the original lattice  at sites $1$ and $3$ there is $A$ and $\emptyset_1$, respectively, which will change
to $A$ and $A$. In the original bases this process can most easily be realized through $A|\emptyset_2|\emptyset_1 \to A|\emptyset_2| A$, which i) has a rate $\lambda_2$ and ii) the effective
rate does not depend on the fact that site $1$ is occupied. Consequently during renormalization new degrees of
freedom will appear and there is no systematic decrease of the energy-scale.

Thus we can conclude that the renormalization group scheme does not work for the generalized contact process. Therefore, it is improbable that a strong disorder fixed point with the scaling properties described in section II.B could be present in this model for some finite value of the disorder. The qualitative difference between the random contact process and the generalized random contact process is due to the competition between the different absorbing states in the latter model. For this case a large death rate indirectly promotes particle branching, since the active phase intrudes between the different absorbing states. This is the reason why the pure model is always in the active phase for $n \ge 3$ \ \cite{hcv01}. This could also be true for the random model.
\section{Numerical investigations}
Here, we discuss the results of two numerical approaches
which allow us to investigate the disordered contact process as a function
of disorder strength. In this way, we will be able to show that the predictions
made in the previous section are consistent with the numerical results provided
the disorder is strong enough. However, we also find that there is a regime
at small to intermediate disorder in which the critical exponents deviate
from the strong disorder ones and indeed seem to vary continuously.
This is evidence for a line of fixed points. However, the numerical
results available do not allow us to decide whether these are ordinary disorder
fixed points (finite $z$) or strong disorder ones (infinite $z$).

\subsection{One-dimensional random contact process}
In our numerical work, we investigated the particular case for
which the rate $\mu_i=1$ and the branching rate $\lambda_i$ is
distributed according to
\be
R(\lambda) = \left[\delta(\lambda - \lambda_+) + \delta(\lambda - \lambda_-)\right]/2
\label{distr}
\ee
with $\lambda_\pm=\exp{\left(A\pm \sqrt{D}\right)}$. The main advantage of this
distribution is that it allows an exact average over disorder to be made,
at least on lattices that are not too large.
For this choice of $R(\lambda)$ the average value of $\ln \lambda$ equals $A$
whereas the variance is $D$. These can therefore be considered as suitable parameters
to measure the activity and the disorder respectively.

\subsubsection{DMRG studies}
The density matrix renormalisation group (DMRG) method is a numerical technique
that was originally introduced to investigate the properties of quantum spin
or fermion systems. The method allows a precise determination of the properties
(energy, magnetisation profiles, correlations, ...)
of the ground state and low lying excitations of such models. The method is
most succesfull in one dimension. Given the formal similarity between quantum
systems and stochastic ones, several groups started to apply this technique
to interacting particle systems in recent years\cite{Enrico,hcv01}. The method is now known
to work well also in these cases though it cannot give as accurate results
as for the spin chains, mainly because at this moment algorithms
to diagonalise non-hermitean matrices are not as well developped
as those for the hermitean case.

In our DMRG work we made calculations for systems with up to $L=24$ sites.
For $L \leq 14$ we were able to perform an exact average over all possible
realisations of the disorder. For the larger systems sizes we considered
typically around $10^4$ disorder realisations.
These calculations were possible for $D \leq 2$. For larger $D$-values,
we encountered numerical difficulties in the DMRG algorithm.

In any finite system, the stationary state of the contact process is the absorbing
state, i.e. the lattice without any particles. In order to study the
absorbing state phase transition within finite systems, we
therefore worked with open boundaries and put the death rate at the most
right site of the lattice, $\mu_L$, equal to zero. This ensures the presence
of particles in the stationary state. It can be expected that for large
enough systems and for sites that are deep in the bulk the effect of
this boundary will be negligible.

Using the DMRG we then firstly calculated the ground state of the
generator (\ref{hamilton}) with open boundaries.
Within this ground state we calculated the density profile, i.e.
$\rho_i=[\langle n_i \rangle]_{\rm av}$. 

In order to determine the location of the critical point and the critical
exponents $x,\ x_s$ and $\nu_\perp$ we investigated in detail the
behaviour of $\rho_s=\rho_1$ (surface density) and $\rho_{L/2}$ (which we
took as our estimate for the bulk density) as a function of the
parameters $A$ en $D$.

A major problem of the contact process in comparison with several well
studied disordered quantum spin chains is that in the present case
the location of the critical point is not known exactly. The
numerical inaccuracy in the location of the critical point will
in its turn influence the accuracy by which critical exponents
can be determined.

To locate the critical point we investigated the quantity
\be
Y_L = \frac{d\ln \frac{d\rho}{dL}}{d\ln L}
\label{defY}
\ee
For a homogeneous system, $\rho$ reaches its large $L$-value
exponentially fast away from the critical point, and as
a power law at the critical point (see (\ref{scaling}) and
(\ref{escaling}) which are the same in the stationary state). As a consequence $Y_L$
goes to $-\infty$ for a non-critical system and to $-(1+x)$ at criticality.
We can therefore expect that in a finite system, the quantity $Y_L$
goes through a maximum as a function of $\mu/\lambda$ from which
one can obtain a finite size estimate of the location of the
critical point and of $x$.
In this way we have obtained the critical
value of $A$, denoted as $A_c$ for different values of the disorder
strength $D$ and for different $L$-values. An extrapolation for
$L \to \infty$ then gives our final estimates for $A_c(D)$.
The data for the surface density $\rho_s$ can be analysed in a completely
similar way and give an independent estimate for the location
of the critical point.
From the analysis in section II, we expect that for $D \to \infty$, the critical
point obeys $[\ln \mu]_{\rm av} = [\ln J]_{\rm av}$ which for the present case
leads to the prediction of the exact location of the critical point, at $A_c=
\ln{\sqrt{8}}=1.0397$. Unfortunately, we are not able to investigate very
large $D$-values and hence could not verify this prediction. 

\begin{figure}
\resizebox{8cm}{8cm}{\includegraphics{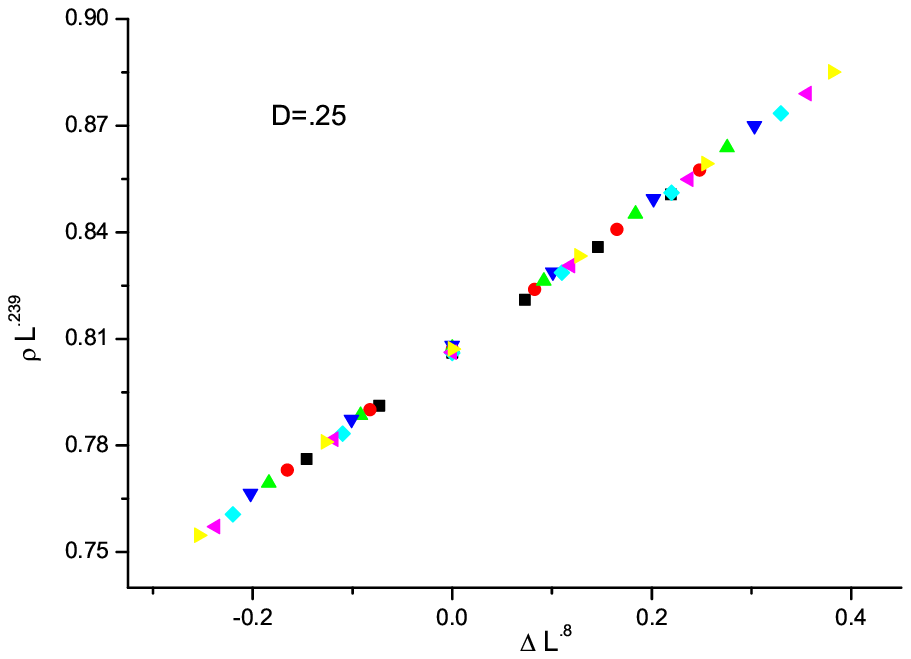}}
\caption{Scaling plot of the particle density at $D=.25$ assuming $A_c=1.19,\ x=.24$
and $\nu_\perp=1.25$. The different symbols indicate $L=12 (\Box), L=14 (\bigcirc),
L=16 (\bigtriangleup), L=18 (\bigtriangledown), L=20 (\Diamond), L=22 (\lhd)$ and
$L=24 (\rhd)$.\label{fig.2}}
\end{figure}

To determine the stationary state critical exponents we next
made scaling plots for the density and the surface density assuming
(\ref{scaling}) (taking $t \to \infty$).
In Fig. \ref{fig.2} we show such a scaling plot for the density and for $D=.25$ and
$A_c=1.19$.
From this we obtain  $x=.24$ and $\nu_\perp=1.25$.
Data for the bulk and surface density for other values of $D$ can be analysed in a
completely similar way and they give rise to the DMRG estimates for the
exponents $x_s$ and $x$  as shown in Fig.3 of reference 31 and Fig. \ref{fig.5} respectively.
From these we see that $x_s$ assumes the value predicted for the strong disorder
fixed point at $D \stackrel{>}{\sim} 1.5 $. For smaller $D$-values, $x_s$ decreases continuously
starting from its value for a homogeneous contact process ($x_s \approx .669$ \ \cite{denen}).

A rather similar behaviour is found for the bulk exponent $x$. For increasing
$D$, its value decreases from that for the homogenous contact process, $x \approx .252$,
to a value of $x \approx .2$ in the region $1.5 \leq D \leq 2$. Here it has
to be remarked that since the DMRG can only be performed for systems with $L$ up
to $24$, it may very well be that our estimate of the bulk density $\rho$ is
still influenced by the boundary conditions which we had to choose, and this
may very well be the reason why the exponent $x$ reaches its value
at the strong disorder fixed point more slowly.

The exponent $\nu_\perp$ is even more difficult to estimate
since its value is very sensitive to the precise location of the critical
point. The values that we have determined are only rough estimates. We find that the value
of $\nu_\perp$ increases from the pure system value with increasing disorder, and
equals $1.67 \pm .08$ at $D=1.5$.

\begin{figure}
\resizebox{8cm}{12cm}{\includegraphics{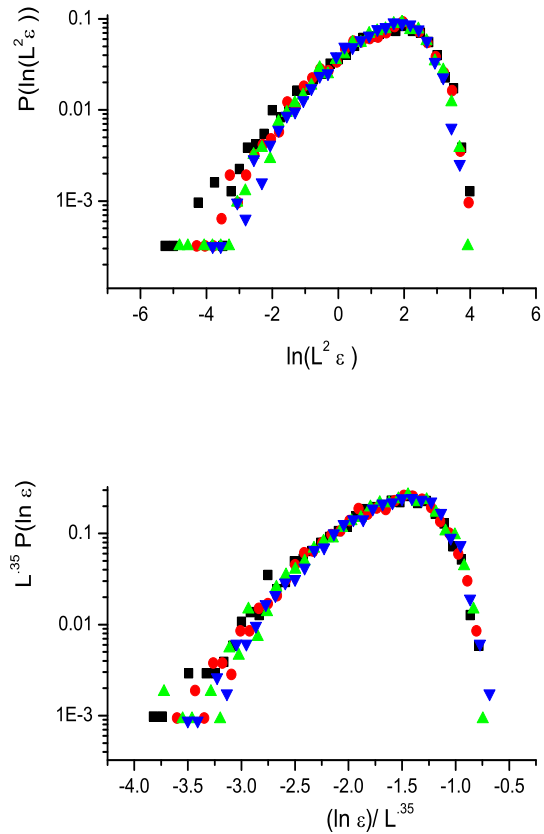}}
\caption{Test of dynamical scaling form for $D=.5, A_c=1.19$. In the upper graph, we
have assumed ordinary dynamical scaling with $z=2.0$, while in the lower graph we used
logarithmic scaling with $\psi=.35$. The different symbols correspond with $L=18 (\bigtriangledown), L=20 (\bigtriangleup), L=22 (\bigcirc)$ and $L=24 (\Box)$.
\label{fig.3extra}}
\end{figure}

The exponent $z$, or in case of a logaritmic scaling $\psi$, can in
principle be determined
from the distribution of the gap in the spectrum of the generator (\ref{hamilton}).
Using the DMRG, we therefore also calculated the distribution of gap sizes for
different system sizes.
In Fig. \ref{fig.3extra} we show the results of such an analysis at the critical point
and for $D=.5$. In the upper graph we show a plot assuming ordinary scaling and the
value $z=2.$, while in the lower part we have assumed logarithmic scaling and
$\psi=.35$. As can be seen from this figure, the DMRG-results do not allow us to
discriminate between the two types of dynamical scaling. A similar conclusion
holds for other $D$-values.

\subsubsection{Monte Carlo simulations}
In an attempt to reach bigger system sizes and to get independent estimates for
the exponents we also performed extensive numerical simulations taking a
single seed particle as initial condition. In order to have a good comparison
with DMRG-data, we again put all death rates equal to one, while
the branching rates are distributed according to (\ref{distr}). For various values of $D$ and
$A$ we calculated the survival probability $P_s(t)$, the total number of particles
in the system
$N(t)$ as well as their average spread $R^2(t)$. We typically simulated up to $t=10^5$ and
took an average over $2.10^4$ disorder realisations. 
For large $D$-values, it is difficult to obtain reliable simulation
data since the dynamics becomes extremely slow. For this reason we are not
able to explore values of $D$ that are much larger then $\approx 1.5$.

\begin{figure}
\resizebox{8cm}{8cm}{\includegraphics{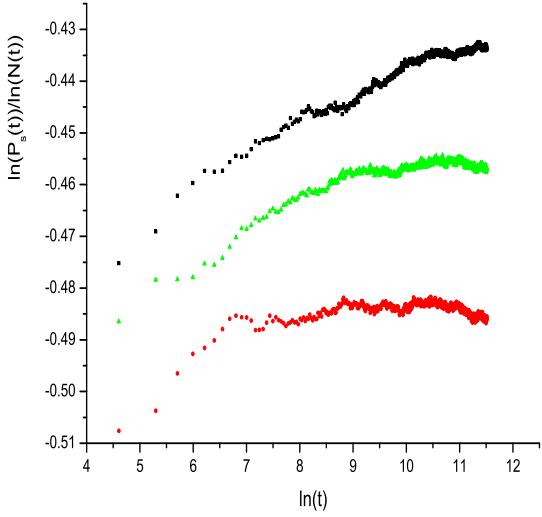}}
\caption{Plots of $\ln P(t)/\ln N(t)$ as a function of $\ln t$ at $D=.5$ and for
$A=1.179,\ 1.177$ and $1.175$ (top to bottom). From this,we estimate
$A_c=1.177$.\label{fig.3}}
\end{figure}

To analyse our data we first have to determine the location of the critical point
$A_c(D)$. While in homogeneous systems the critical point is the only one
characterised by a power law decay of $P_s(t)$, the same is not true
in a disordered system as first pointed out by Bramson, Durrett and Schonmann \cite{BramDur}. Power law behaviour
can indeed be found in the whole subcritical regime and is a manifestation
of the presence of a Griffiths phase. Still, we expect $N(t)$ to increase above
$A_c$ and to decrease below criticality. In this way we obtain a first
estimate of the location of the critical point.
A second criterion which we used is that both for convential and
for strong disorder scaling, at criticality, $\ln{P}/\ln{N}$ becomes
a constant asymptotically. In Fig. \ref{fig.3} we show some typical set of data for
this quantity taken
at $D=0.5$. From these we determine $A_c(0.5)=1.177$. The values for the critical activities
which we find in this way are close (approximately within $1 \%$) to these found from the DMRG.

\begin{figure}
\resizebox{8cm}{12cm}{\includegraphics{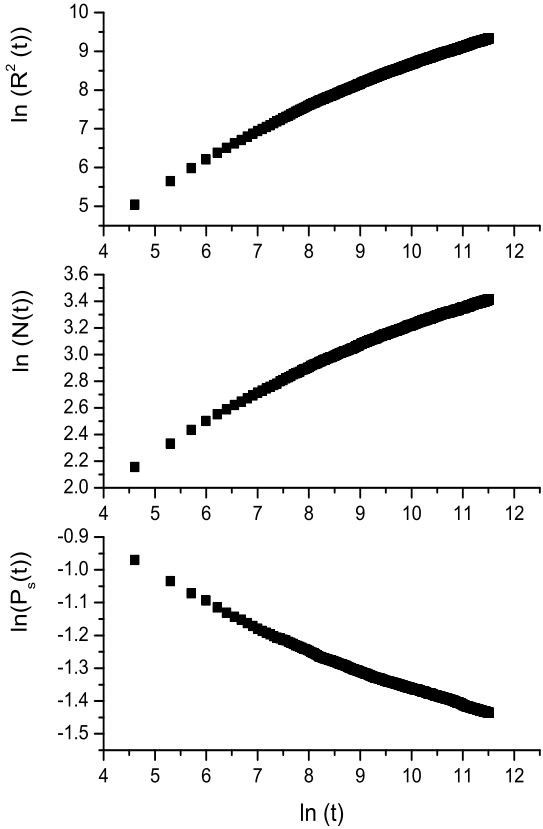}}
\caption{Plot of the survival probability $P_s(t)$, the number of particles
$N(t)$ and their average spread $R^2(t)$ (bottom to top) as a function of
$t$ for $D=1, A_c=1.171$.\label{fig.4}}
\end{figure}

In Fig. \ref{fig.4} we present log-log plots for the three quantities of interest
at the critical point with $D=1$. These results are typical also for the other
$D$-values investigated.

\begin{figure}
\resizebox{8cm}{8cm}{\includegraphics{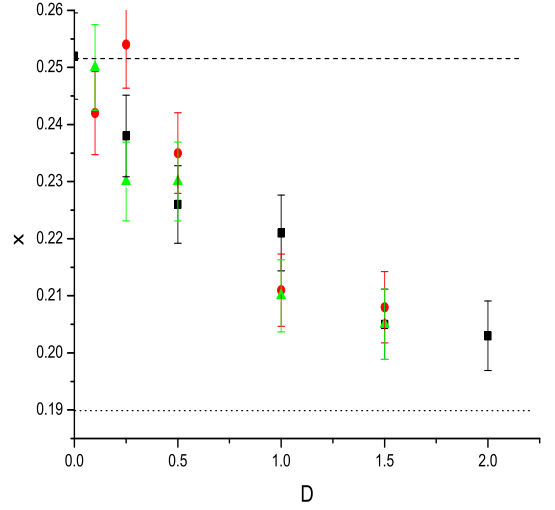}}
\caption{Plot of the exponent $x$ versus disorder strength $D$. The different symbols indicate
the results obtained from the DMRG (squares), and the simulations
assuming ordinary scaling (circles) and logarithmic scaling (triangles).
We also indicate the values for strong disorder (dotted line) and for the
homogeneous system (dashed line).\label{fig.5}}
\end{figure}

\begin{figure}
\resizebox{8cm}{8cm}{\includegraphics{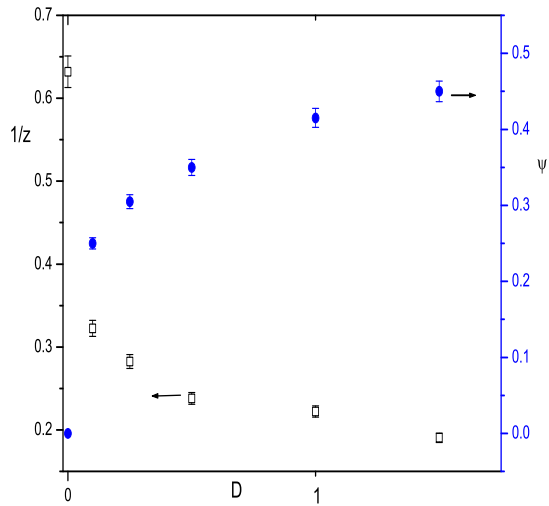}}
\caption{Plot of the exponents $1/z$(squares) and $\psi$(circles) as a function of disorder
strength $D$.\label{fig.6}}
\end{figure}

As can be seen there is still some curvature visible in these figures, which could
e.g. arise from the logarithmic corrections which are ubiquitously present in these
kind of random 'quantum' systems. Yet, the late time part can be fitted quite well
to a power law from which we can determine estimates of the exponents $x$ and $z$ as a
function of $D$. Our results for these exponents are presented in Fig. \ref{fig.5} and Fig. \ref{fig.6}
respectively.

We notice
two interesting trends. Firstly, we observe a decrease of $x$ from its homogenous
system value to a value that is consistent with that at the infinite disorder
fixed point. The numerical values are moreover consistent with those found
from the DMRG. Secondly, we observe that the dynamical exponent $z$ seems to
make a jump as soon as any disorder is present after which it increases with
$D$. These results are then consistent with the idea of a line of ordinary
(disorder) critical points which ends at some critical value of the disorder
above which exponents assume precisely the values of the strong disorder
fixed point.

In a regime governed by strong disorder fixed points we expect that the dynamical
quantities do not follow a power law but scale logarithmically, i.e. as given in (\ref{logexp}).
Indeed such a behaviour was found in the simulations in $d=2$ \ \cite{weerDick}.
We found that this kind of scaling can also describe our results 
for all $D$-values investigated. Assuming strong disorder
scaling, we can determine estimates for the exponents $x$ and $\psi$.
These values are also shown in Fig. \ref{fig.5} and Fig. \ref{fig.6}. We notice that the values for the exponent $x$ are
not very sensitive to the type of scaling that we assumed. We also observe
that the exponent $\psi$ seems to increase towards its value expected at the
strong disorder fixed point, i.e. $\psi=1/2$.

We can thus see that, as was for the case for the DMRG, it is not possible to rule
out from the numerics that there is a line of strong disorder fixed points in the
model.
When we compare the estimates for the dynamical exponents coming from
the two numerical approaches, we find that those for the logarithmic corrections
are more selfconsistent.
There is a rather large discrepancy between the $z$-values as found from the DMRG
and the simulations. This could be due to a lack of asymptoticity in time for the
simulations and in size for the DMRG. Assuming logarithmic scaling, the values
which we find from the two approaches are almost the same. 
 The existence of a line of strong disorder fixed
points was not found so far in real quantum systems and could
be an essential new feature of stochastic models. Yet, at present we have no theoretical
underpinning for the existence of such a line of disorder fixed points.
Further numerical studies are needed to obtain more insight on this point.
 
\subsection{Two-dimensional random contact process}
\begin{figure}
\resizebox{8cm}{8cm}{\includegraphics{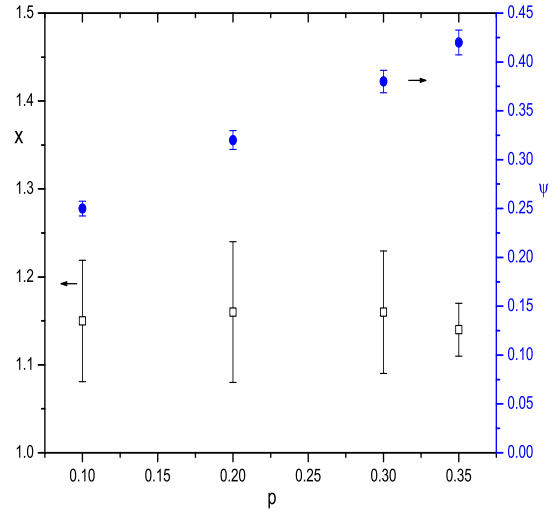}}
\caption{Values for the exponents $x$(squares) and $\psi$(circles) as a function of dilution $p$ assuming logarithmic scaling. The values in this table are for $d=2$ and are determined from the numbers given
in table I of the first paper in reference 15.}
\label{fig.7}
\end{figure}

As already remarked above, in the second paper of reference 15, the logarithmic
scaling (\ref{logexp}) was first observed in the two-dimensional contact process
with dilution where it was dubbed 'violation of scaling'. In the present work
we have shown that this kind of scaling is the natural one to be expected at
a strong disorder fixed point. In reference 15 it was assumed that the
logarithmic scaling holds for all values of the dilution, which here we will
denote by $p$. Since we don't have the original dataset, it is not possible
to investigate whether the data of these authors are also consistent with
ordinary scaling in the small $p$-regime. From the numerically determined
values of $\overline{\theta},\ \overline{\eta}$ and $\overline{\sigma}$ it is
however possible to determine values for the exponents $x$ and $\psi$ in the
two-dimensional case. The results are shown in Fig. \ref{fig.7}.

These numbers can now be compared with those expected to hold at the strong disorder
fixed point of the RTIM in two dimensions, which are $x=1.0$ and $\psi=.42$.
Note that for the largest disorder value, the exponent $\psi$ is very close
to this value. The exponent $x$ seems larger then expected.
However, in this respect it is interesting to remark that the authors of
ref. 15 notice that the data on $N(t)$ are consistent with ordinary
scaling but with $\eta=0$ or with logarithmic scaling and a very small
value of $\overline{\eta}$. This is consistent with our ideas which predict
that at the strong disorder fixed point in $d=2$ and for $x=1$, $\overline{\eta}$
equals zero.
Hence, we believe that also in two dimensions the strongly disordered
contact process is in the same universality class as the RTIM.

\section{Conclusions}
In this paper, we have investigated the effect of quenched disorder in the transition
rates on the critical behaviour of some models with absorbing state phase
transitions.
For models in the directed percolation universality class, we have given
convincing evidence that, if the disorder is large enough, the
universal behaviour is that known from the RTIM and some other disordered
quantum spin chains. This result follows from calculations using a strong
disorder renormalisation approach and is consistent with numerical results
in one dimension
obtained with the DMRG and simulations.
A reanalysis of data for a diluted contacted process in two dimensions
is also consistent with this conclusion.
It therefore seems that the effects of strong disorder lead to a new kind
of universality between hermitean and non-hermitean quantum models.

For weak and intermediate disorder our numerical results indicate the existence
of a line of fixed points with continuously varying exponents. Such behaviour
has been found previously in some quantum spin chains \cite{EnricoFerenc}. 
Most interestingly, our data could be consistent with a scenario in which
this is a line of strong disorder fixed points. Such a behaviour is also
consistent with data obtained by Moreira and Dickman on a two-dimensional
diluted contact process.
Further simulations are needed to confirm this picture. It would be especially
interesting to develop techniques that can efficiently simulate these
kind of systems for sufficiently large disorder and long enough times.
Moreover, it would be very interesting to investigate whether the behaviour
that we find can also be seen for distributions of the transition rates
different from the ones that we used in our simulations.
If the logarithmic dynamical behaviour can be demonstrated, it becomes
a challenge to understand such behaviour from analytical or RG arguments.

For absorbing state phase transitions in other universality classes we have
fewer results. Our RG calculations indicate that for the generalised
contact process with disorder, it may be possible that the model is
always active, also for $n=2$. This prediction should be verifiable
with simulations. Whether this result also holds for other models
in the same universality class is not clear yet. We believe that
the effect of quenched disorder on absorbing state phase transitions in particular,
and on stochastic many-particle systems in general, provides an interesting
field of future research.

\appendix


\section{Decimation for the random contact process}

When the branching rate between two sites (say $2$ and $3$) is the largest rate in the
system, we consider the Hamiltonian
(\ref{hamilton}) restricted to these two sites and with free boundary conditions.
This operator can be represented by the
matrix:
\be
{\bf H_{CP}^{23}} = \left(
\matrix{
 0  &     -\mu_3        &      -\mu_2       &       0      \cr
 0  & \mu_3+\lambda_2 &         0         &    -\mu_2    \cr
 0  &         0         & \mu_2+\lambda_2 &    -\mu_3    \cr
 0  &   -\lambda_2    &   -\lambda_2    &  \mu_2+\mu_3 \cr
}
\right).
\label{h23}
\ee
In the limit $\mu_2/\lambda_2 \to 0$ and $\mu_3/\lambda_2 \to 0$ the eigenvalues are $0,0,\lambda_2,\lambda_2$. The degeneracies disappear for finite $\mu_2/\lambda$
and $\mu_3/\lambda$. Discarding the two highest energy levels, the gap between the remaining two
lowest states can be calculated perturbatively leading to:
\be
E_{CP}^{23}=\frac{2\mu_2 \mu_3}{\lambda_2}.
\label{e23}
\ee
Now keeping in mind that a one-site cluster with a death rate $\tilde{\mu}$ has
a gap $\tilde{\mu}$ we obtain for the renormalized death rate,
$\tilde{\mu}_2=E_{CP}^{23}$, as announced in (\ref{mutilde}).

When the largest rate is a death rate, say $\mu_2$, one considers a three site cluster whose Hamiltonian (free boundary conditions) is represented by the matrix:
\begin{widetext}
\be
{\bf H_{CP}^{123}} = \left(
\matrix{
 0  &     -\mu_2  &  0  & 0  &  0  & 0 &  0  & 0   \cr
 0  & \mu_2+\lambda_2+\lambda_3 & 0  & 0  &  0  & 0 &  0  & 0  \cr
 0  & 0 & \lambda_2 & -\mu_2 &  0  & 0 &  0  & 0  \cr
 0  &   -\lambda_2    &   -\lambda_2    &  \mu_2+\lambda_3 &  0  & 0 &  0  & 0  \cr
 0  & 0 &  0  & 0 & \lambda_3 & -\mu_2 &  0  & 0 \cr
 0  &   -\lambda_3  & 0  & 0 & -\lambda_3    &  \mu_2+\lambda_2 &  0  & 0 \cr
 0  & 0  &  0  & 0 &  0  & 0 & \lambda_2+\lambda_3 & -\mu_2 \cr
 0  & 0 &  0  & -\lambda_3 &  0  & -\lambda_2 & -\lambda_2-\lambda_3 & \mu_2 \cr
}
\right).
\label{h123}
\ee
\end{widetext}
The corresponding eigenvalue problem is reduced to the diagonalization of four $2\times 2$
matrices. In each subspace there is an eigenvalue of $O(\mu_2)$, which is discarded. Of the remaining four lowest eigenvalues there are two zero and two with
the value
\be
E_{CP}^{123}=\frac{\lambda_2 \lambda_3}{\mu_2}\;.
\label{e123}
\ee
If we keep in mind that in a two-site cluster with an effective branching rate $\tilde{\lambda}$
the spectrum is given by $0,0,\tilde{\lambda},\tilde{\lambda}$ we obtain for the renormalized rate
$\tilde{\lambda}=E_{CP}^{123}$, the result stated in (\ref{lambdatilde}).

\section{Solution of the RG equations at the critical point}

We look for a special
solution of (\ref{Rdiff}) and (\ref{Pdiff}) (in the fixed point $\Omega \to 0$)
of the form:
\begin{eqnarray}
R(J,\Omega)=R(\Omega,\Omega)\left(\frac{\Omega}{J}\right)^{1-R(\Omega,\Omega)\Omega}
\label{Rsol}\\
P(\mu,\Omega)=P(\Omega,\Omega)\left(\frac{\Omega}{\mu}\right)^{1-P(\Omega,\Omega)\Omega}
\;.
\label{Psol}
\end{eqnarray}
Substituting (\ref{Rsol}) and (\ref{Psol}) into (\ref{Rdiff})
we get as:
\begin{equation}
\frac{{\rm d} \ln \tilde{R}}{{\rm d} \Omega}-\frac{{\rm d} \Omega \tilde{R}}{{\rm d} \Omega}
\ln \frac{\Omega}{J} + \frac{1-\Omega \tilde{R}}{\Omega}=\tilde{P}-\tilde{R}
-\Omega \tilde{R} \tilde{P} \ln \frac{\Omega \kappa}{J} \kappa^{-\Omega \tilde{R}}\;,
\label{Rint}
\end{equation}
where we used the notations, $\tilde{P}(\Omega) \equiv
P(\Omega,\Omega)$ and $\tilde{R}(\Omega) \equiv R(\Omega,\Omega)$.
For a moment we set $\kappa=1$. After a trivial rearrangement, we obtain the relation:
\begin{equation}
\left[\Omega \tilde{R} \tilde{P}-\frac{{\rm d} \Omega \tilde{R}}{{\rm d} \Omega}
\right]\left[\ln \frac{\Omega}{J}-\frac{1}{\Omega \tilde{R}} \right]=0
\;,
\label{Rint1}
\end{equation}
which leads to the ordinary differential equations:
\begin{eqnarray}
\frac{{\rm d} \tilde{R}}{{\rm d}\Omega}=-\frac{\tilde{R}}{\Omega}+\tilde{P}\tilde{R}
\label{Redge}\\
\frac{{\rm d} \tilde{P}}{{\rm d}\Omega}=-\frac{\tilde{P}}{\Omega}+\tilde{P}\tilde{R}
\;.
\label{Pedge}
\end{eqnarray}

These equations, which hold just for $\kappa=1$ can be solved for general (non-symmetric) distributions\cite{i02}. The solution gives information about the singular behavior of the system outside the critical point, i.e. in the Griffiths phase. This general solution, however, does not apply for $\kappa \ne 1$. In contrary, the solution at the fixed point, when $R(J,\Omega)$ and $P(\mu,\Omega)$ are identical, thus $\tilde{R}= \tilde{P}$, holds for any finite $\kappa > 0$. Indeed, in terms of the variable, $y=\tilde{R} \Omega=\tilde{P}\Omega$, and the log-energy scale, $\Gamma=-\ln \Omega$ we obtain the differential equation:
\begin{equation}
\frac{{\rm d} y}{{\rm d} \Gamma} + y^2=0,
\;,
\label{dif3}
\end{equation}
with solution:
\begin{equation}
y=\tilde{R} \Omega=\tilde{P}\Omega=\frac{1}{\Gamma-\Gamma_0}=\frac{1}{\ln(\Omega_0/\Omega)},
\quad \delta=0
\;,
\label{sol0}
\end{equation}
Here $\Gamma_0=-\ln \Omega_0$ is a reference (log)energy scale. Now, going back to (\ref{Rint}) we can see, that at the fixed point the actual value of $\kappa$ does not matter. The term $\kappa^{-\Omega \tilde{R}} \to 1$ and $\ln \kappa / \ln (\Omega/J) \sim
\kappa \tilde{R} \Omega \to 0$.
\\ 
\section{The Random transverse-field Ising model and its decimation rules}

The prototype of random quantum systems is the
random transverse-field Ising model which is defined by the Hamiltonian:
\be
H=-\sum_{\langle i,j \rangle} J_{ij} \sigma_i^x \sigma_j^x-\sum_i h_i \sigma_i^z\;.
\label{hamilton2}
\ee
Here the sum runs over nearest neighbors and $\sigma_i^x$, $\sigma_i^z$ are Pauli matrices at site $i$. The
 exchange couplings $J_{ij}$ and the  transverse-fields $h_i$ are independently distributed
random variables with distributions $\pi(J)$ and $\rho(h)$,
respectively. The Hamiltonian in Eq.(\ref{hamilton2}) in one dimension is closely related
to the transfer matrix of a classical two-dimensional layered Ising
model, which was first introduced and studied by McCoy and Wu \cite{mccoywu}.

In the renormalization scheme we have the following decimation relations. If the largest term in the Hamiltonian is a coupling, say $J=\Omega$, then the two sites connected by $J$ and having transverse fields, $h$ and $h'$ and moments $m$ and $m'$, behave as an effective composite cluster with moment $\tilde{m}$ in a renormalized transverse field, $\tilde{h}$, which are given by:
\be
\tilde{h} = \frac{h h' }{J},
\quad \tilde{m}=m+m'\;.
\label{h_deci}
\ee
On the other hand if the largest term is a transverse field, say $h=\Omega$, then the site with this transverse field gives negligible contribution to the (longitudinal) magnetic susceptibility, thus can be decimated out. This leads to a renormalised coupling between the nearest neighbours
of the decimated site that is given by:
\be
\tilde{J} = \frac{J J' }{h},
\quad \tilde{m}=m+m'\;.
\label{J_deci}
\ee
which is related to (\ref{h_deci}) through duality.

\ \\

{\bf Acknowledgment}: J.H. is thankful to G.T. Barkema for useful disucussions concerning the
simulations. F.I. is grateful to P. Grassberger, J.D. Noh and F. van Wijland for
stimulating discussions and to L. Gr\'an\'asy for his help in the numerical
calculations. F.I.'s  work has been supported by the Hungarian National
Research Fund under  grant No OTKA TO34183, TO37323,
MO28418 and M36803, by the Ministry of Education under grant No FKFP 87/2001,
by the EC Centre of Excellence (No. ICA1-CT-2000-70029) and the numerical
calculations by NIIF 1030. 
In the early stages of this work, J.H. was supported by the FWO-Vlaanderen.


\begin{thebibliography}{99}
\bibitem{note_gr} This relation follows from the fact that in a finite system of size, $l_c$, the smallest gap in the active phase behaves as $\epsilon \sim \exp(-\sigma l_c)$, thus the ground state is degenerate in the thermodynamic limit.
\bibitem{note1}
 Here generally the parameter, $\kappa$, should satisfy $\kappa \ge 1$, unless the resulting fixed-point distribution is singular, so that the integral with the
analytical continuation of $R(J,\Omega)$,
$\int_{\Omega}^{\kappa \Omega} R(J,\Omega) {\rm d} J \to 0$, for any $\kappa>0$. We shall see that the fixed-point distribution indeed satisfies this condition.
\end{thebibliography}
\end{document}